# Towards Green AI: Current Status and Future Research

Christian Clemm
*Dept. of Precision Engineering*
The University of Tokyo
Tokyo, Japan

Lutz Stobbe
*Dept. Environmental and Reliability Engineering*
Fraunhofer Institute for Reliability and Microintegration IZM
Berlin, Germany

Kishan Wimalawarne
*Dept. of Mathematical Informatics*
The University of Tokyo
Tokyo, Japan

Jan Druschke
*Dept. Environmental and Reliability Engineering*
Fraunhofer Institute for Reliability and Microintegration IZM
Berlin, Germany

*Abstract*— The immense technological progress in artificial intelligence research and applications is increasingly drawing attention to the environmental sustainability of such systems, a field that has been termed 'Green AI'. With this contribution we aim to broaden the discourse on Green AI by investigating the current status of approaches to both environmental assessment and ecodesign of AI systems. We propose a life-cycle-based system thinking approach that accounts for the four key elements of these software-hardware-systems: model, data, server, and cloud. We conduct an exemplary estimation of the carbon footprint of relevant compute hardware and highlight the need to further investigate methods for Green AI and ways to facilitate wide-spread adoption of its principles. We envision that AI could be leveraged to mitigate its own environmental challenges, which we denote as 'AI4greenAI'.

## I. Introduction

We are in the midst of an explosive growth of the development and integration of artificial intelligence (AI)-based systems into all aspects of human activities that has been speculated to be 'as transformative as the industrial revolution' and could incur profound social and economic changes [1]. The release of 'generative AI' applications, notably the text generator ChatGPT, text-to-image generators like Midjourney, and text-to-video models like Sora have recently brought public attention to the rapidly progressing technological capabilities. AI has been poised to be capable of accelerating breakthroughs in science [2], medicine [3], and other fields, due to increasingly sophisticated algorithms and advancing availability of the required computational power.

However, besides the recognition of the immense potential of AI systems, the rapid progress has raised concerns about their potential impacts on societies, shared by researchers, policy-makers, and the public [4]. Besides social impacts, such as the expected transformation of the job market, the environmental consequences of the broad deployment of resource- and energy-intensive hardware-software systems raise concerns. The trend towards rapidly growing model complexity and size, trained on expansive volumes of data, has been suggested to be a prerequisite for increasing the performance of AI models. Reported 'neural scaling laws' [5] are based on the empirical observation of power-laws that govern the scaling of neural network performance with model and dataset size [6]. Accordingly, concurrent increases in model parameters, training tokens, and computational power are required to uphold the trend towards increasing model performance, in some ways reminiscent of Moore's law, which has been driving the miniaturization and performance progression of semiconductor technologies for decades [7].

Indeed, assessing the growth in model complexity and consumed energy for the training of OpenAI's generative pre-trained transformer (GPT) models, the large language model (LLM) behind the popular ChatGPT, reveals that the model size has increased from 1.5 billion to an estimated 1.7 trillion parameters between GPT-2 and GPT-4, an increase by a factor of 1,000 [8][9]. Based on approximations, the energy consumed for the training of GPT models has grown by a factor of 2,000 between the release of GPT-2 and GPT-4 in a time span of just 3.5 years [8][10], as is displayed in Fig. 1. To provide context, the energy used to train and run ChatGPT has been estimated to be equal to the annual carbon emissions of 175,000 Danish citizens [11].

The rapidly growing computational requirements of AI models necessitate increasingly powerful hardware to provide the computational infrastructure required for the training and inference of AI models. Graphics processing units (GPU) provide the parallel processing capabilities and are employed in server systems operated in globally distributed data centers ('the cloud'). The energy needs of the compute hardware and required heating, ventilation, and air conditioning (HVAC) in data centers are ever-increasing. The IEA projects the electricity consumption of data centers could reach 1,000 TWh in 2026, roughly equivalent to the electricity consumption of Japan [12]. Some have voiced the idea to power AI-centric data centers via built-in nuclear reactors [13]. Besides energy, the resources needed to fabricate the integrated circuits and periphery, including critical raw materials (CRM), and to operate data centers, also require attention.

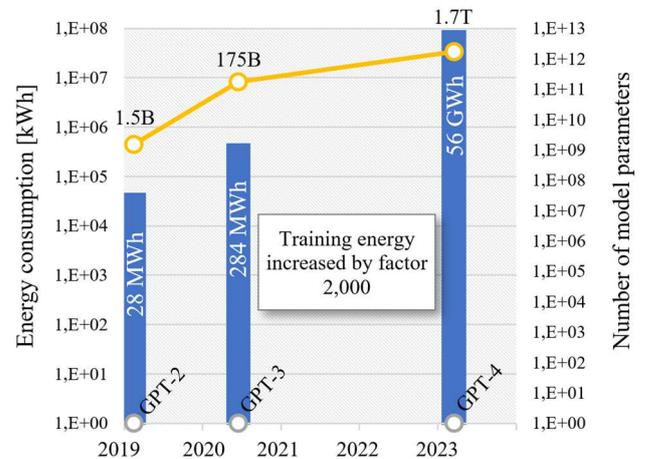

Fig. 1. Growth in GPT model size and training energy consumption

These concerns have led to increasing interest in 'Green AI', a term denoting "AI research that yields novel results while taking into account the computational cost" [14]. The goal of this paper is to broaden the scientific discourse on Green AI. We discuss the status and future research needed to support the development of Green AI through the lens of system- and life-cycle-thinking accounting for both 'assessment' and 'ecodesign' of the underlying software and hardware systems. We aim to identify limitations of the current status and discuss a vision of future research directions. In doing so, we do not set our focus primarily on 'green by AI', the notion of leveraging AI to solve sustainability challenges, such as training models to steer systems towards higher efficiency.

In this work, we use the terms 'green' and 'sustainable' synonymously with the environmental dimension of sustainability without implying the social and economic dimensions.

The remainder of this paper is organized as follows: Section II provides background information on the basics of AI, Green AI, and environmental assessment of such systems. Section III introduces an analysis framework that supports system- and life-cycle-thinking for environmental assessments and optimization of AI systems. Section IV provides an overview on the current status of the application of life cycle assessment methods to AI systems and provides illustrative examples. Section V discusses approaches and techniques that have been proposed to enhance the environmental compatibility of AI systems. Section VI discusses future research that is needed to support the further development and broad uptake of Green AI paradigms.

## II. BACKGROUND

This section summarizes basic concepts of AI systems, Green AI, and of environmental impact pathways and assessment methods applied to AI systems.

### A. Background on AI Systems

AI is the overarching term for the science of artificial intelligence, in which computers are used to simulate intelligent behavior, such as learning, judgment, and decision-making. Machine learning (ML) is a subfield of AI focused algorithms that are capable of learning from data [15]. ML models are trained on data to automatically identify patterns, which can then be used to make predictions or decisions on previously unseen data [16], known as 'inference'. Deep learning (DL) is a subset of ML and refers to an approach that utilizes artificial neural networks that are composed of layers of stacked artificial neurons employed to approximate complex, non-linear functions [17]. DL is at the core of the dramatic improvements in the state-of-the-art in speech recognition, visual object recognition, object detection and many other domains [18]. DL is used for a wide number of tasks such as classification, regression, and generative modelling. A common example is image recognition, where the input is an image file and the output is a probability distribution among numerous potential objects (Fig. 2, adapted from [17]). Other popular applications include so-called 'generative AI', including large language models (LLM) for chatbots, image, voice, music, and video generators.

Model size is often identified via its number of parameters, which are coefficients in the model that are adjusted during the training process. The size of the training dataset is often identified via the number of features (e.g., tokens of a language model, pixels of an image, nodes of a graph) and number of entries. The performance of models is, inter alia, governed by hyperparameters that are set before training, such as the number of nodes and layers of a neural network. To find the best set of hyperparameters, models are typically trained repeatedly with different sets of hyperparameters, multiplying training time, cost, and energy consumption.

An AI model, at its core, is software code that is executed to perform calculations on specialized hardware that is run on servers, commonly located in data centers ('the cloud'). Most of the environmental impacts of AI systems derive from the manufacturing, installation, operation, and end-of-life treatment of the server systems and supporting data center and communication network infrastructure. The hardware for AI is typically based on an x86-type CPU (central processing unit), GPU (graphics processing unit) or TPU (tensor processing unit) that are operated as part of a server. CPU systems are based on software and memory (DRAM) which enables relatively slow sequential computing. CPUs are typically utilized for small AI models which require relatively short training cycles. GPUs are based on arithmetic logic units (ALU) and allow parallel computing to a much higher degree in comparison to CPUs. GPUs are therefore faster and allow calculations of medium to large AI models with considerable batch sizes. GPUs require considerable memory capacity which is typically provided through high bandwidth memory (HBM). TPUs are application-specific processors consisting of matrix multiplication units (MUX) in a systolic array computer architecture which work in conjunction with HBM. Memory access during matrix multiplication is not necessary which enables exceedingly high speeds. Both GPUs and TPUs are used for matrix calculations and large AI models with considerable complex training phases.

Data-driven processes (data mining), algorithms (ML), or artificial neural networks (DL) as well as big data analytics characterize AI systems. They require big data systems with specific properties to be capable of learning, modeling, and model application. These systems are necessarily distributed, as the data has become too large to fit on a single machine, be it hard drives or random access memory (RAM). Therefore, any algorithm used to analyze big data systems must be parallelizable. The data is available in different and unstructured formats (e.g., audio, text, or video) and cannot be handled according to a standardized scheme. In addition, the data is updated very quickly or delivered at a speed that does not allow the entire stream to be stored. The handling of such data streams imposes restrictions in terms of algorithms and fault tolerance. Fault tolerance in big data systems means that the systems continue to work even if individual nodes or network connections fail. This is also an important prerequisite for horizontal scaling, as a system can grow by adding nodes/connections. More compute nodes provide more storage and processing capacity for big data. The system's application programming interface (API) ensures that the application is not affected by the expansion of the system (no reprogramming). The data must be divided and distributed to different nodes, improving the overall application performance. In comparison to traditional data systems, the functionality moves from system level to application level. The reasons are the limitations of relational database systems, which are not good in data distribution and partitioning.

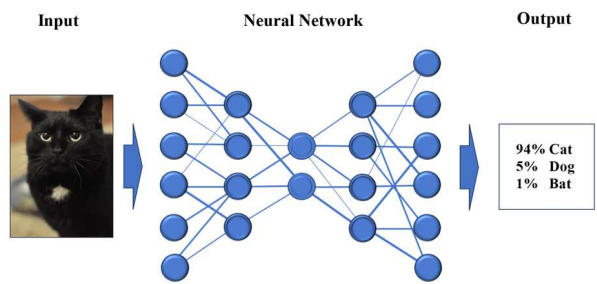

Fig. 2. Schematic illustration of a deep learning neural network

*B. Background on Green AI*

The negative environmental impact of ML is usually directly related to the required high computational power, i.e., the amount of energy required to train and run ML models [19]. While AI research and development has traditionally aimed to maximize model performance, such as minimizing loss functions, Green AI takes account of the environmental footprint and focuses on minimizing computation while still producing accurate results [20][14]. Factors beyond software-related computational power like hardware manufacturing, transportation, or e-waste are harder to quantify and often not considered in literature to date.

A systematic review of literature on Green AI identified 98 studies, with 76% having been published since 2020, indicating the recently growing interest in this field [21]. Among others, the review concluded that most existing literature has addressed Green AI at the level of energy efficiency only [21]. Energy efficiency is related to the computational power required for training and operation of AI systems. Further, most studies in the field thus far have focused on the model training phase, which is frequently highlighted for its immense compute requirements and energy consumption for the training of increasingly complex models [21]. The most frequently addressed topic in Green AI papers was reported to be monitoring approaches to study the energy and/or carbon footprint of AI models, followed by research on hyperparameter tuning, model benchmarking, deployment, and model comparison [21].

In our view, Green AI should be considered to encompass any method and measure that aims to minimize or avoid any category of environmental impact that may arise during the AI/ML system's life cycle, including embodied and operational carbon emissions, material efficiency and circularity, and potentially other relevant categories such as the water footprint.

III. PROPOSED ANALYSIS FRAMEWORK

AI is a complex system of algorithms, model architecture, data sourcing, training, and hardware, among other components, and achieving Green AI requires effort that targets all stages of an AI system's life cycle [22]. We agree with this understanding and propose a framework consisting of the key elements of AI systems on both software and hardware level (Fig. 3).

The software life cycle includes the model and the data. The term "model" represents algorithms and model development. Model development is understood to be continuous and consists of training, deployment, inference, and maintenance. The term "data" represents the big data used for training and validation of the model. This data needs to be identified, sourced, processed, and stored.

The hardware life cycle includes the cloud and the server. The term 'cloud' stands for data centers and networks that provide the IT infrastructure for the AI/ML service. The cloud is not abstract, it has a location and time stamp, which varies the environmental impact considerably. For instance, the carbon intensity of electricity in a particular national energy grid may vary according to the amount of renewable energy sources that are utilized. A large amount of renewable energy in the energy mix will reduce the carbon footprint of the AI/ML system in operation. The second term on the hardware side is servers. The term "servers" represents computer hardware including processors, memory, data storage, and network elements. It also includes the power supply and cooling overheads. This hardware creates a constant environmental impact. It starts with the raw material procurement, manufacturing of the system components, the power consumption in the use phase and the type of end-of-life treatment (EoLT).

We believe that this parallel life cycle thinking approach, including both software and hardware, should build the goal and scope of assessing the environmental impact of AI/ML services. Furthermore, the approach builds the decision-making foundation for ecodesign with the aim of reducing the AI/ML system's environmental impact.

IV. ENVIRONMENTAL ASSESSMENT OF AI SYSTEMS

Discourse on the sustainability of AI systems and the development and application of effective measures to mitigate environmental impacts requires suitable metrics and indicators capable of capturing the full scale of system-wide effects. This section explores methodical approaches and related aspects for environmental assessments of AI/ML systems, following the parallel life cycle thinking approach outlined in section III.

Life cycle assessment (LCA) is an ISO-standardized method used to quantify environmental impacts of products and services (ISO 14040/44). All material and energy inputs and outputs as well as products, waste, and emissions are accounted for along all processes associated with the entire life cycle of a product or service. Environmental impacts are quantified along various categories, including global warming potential (GWP), abiotic resource depletion potential (ADP), ozone depletion potential (ODP), or ecotoxicity and human toxicity potential. Carbon footprinting is a subset of LCA in which the carbon-equivalent emissions of greenhouse gasses are considered as the only impact category. The product carbon footprint (PFC) is a suitable indicator for the environmental impact of the two most important life cycle phases; the manufacturing phase, including raw material extraction, and the use phase, which relates to the power consumption of hardware systems during usage.

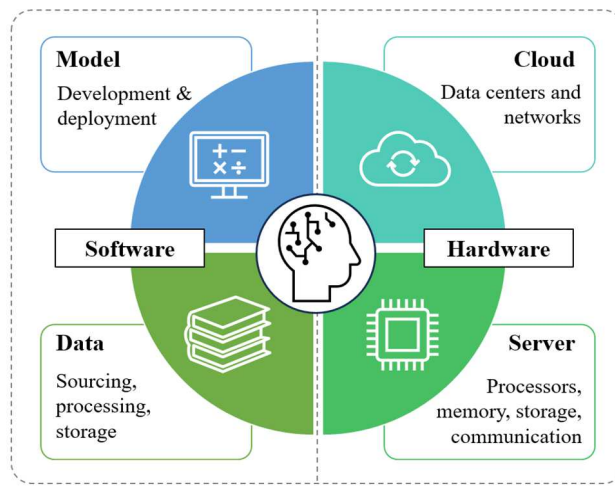

Fig. 3. Proposed framework for the analysis of AI systems

LCA and PFC consider the entire life cycle of products. For physical products, the main life cycle stages are typically raw material sourcing, processing, manufacturing, distribution, usage, and end-of-life treatment (EoLT). In the life cycle of software, such as AI/ML models, the equivalent stages consist of model development and data sourcing, model training, deployment, and decommissioning. In hardware-software systems, the parallel life cycles of hardware and software intersect at the use stage (Fig. 4).

A key decision when applying LCA or PCF is on the functional unit (FU), which is a measure of the function or utility of a product or service to which all environmental effects are scaled. For instance, for a car, a typical FU is a vehicle kilometer traveled. All environmental impacts occurring throughout the car's life cycle, including manufacturing and EoLT, are proportionately allocated to the functional unit.

### A. Software Life Cycle

Unlike physical products, software or code is a virtual good that exists and functions only in conjunction with computer hardware. Environmental assessments of AI systems are therefore largely assessments of the hardware on which the software operates. An environmental assessment should consider the entire hardware system required to create and operate an AI service, including the computer hardware and network components that connect the service to end-users.

The decisions taken during the development of AI/ML models directly determine the computational power required for training and inference, and therefore also the scale of the environmental impacts from hardware manufacturing and usage. Computing power is typically measured in floating-point operations per second (FLOPS), processor (e.g., GPU, CPU) utilization rate, and compute hours [23]. During use, the utilization rate, lifetime, and maintenance of AI services are further variables determining the total energy consumption.

A number of previous studies have evaluated the energy consumption and carbon emissions of AI/ML models. Prominently, reference [24] analyzed the carbon impact of training their own state-of-the-art models, estimating that training one large transformer model takes 274,120 GPU hours, consuming 656,347 kWh energy in the process, resulting in 284 metric tons $CO_2e$ emissions, based on the average U.S. electricity carbon intensity. A further 29 papers that address the carbon footprint or ecological footprint have been identified in a systematic review of Green AI literature [21]. From the previous literature, it can be concluded that the training phase of AI/ML models has thus far been the largest factor in life cycle energy consumption. The training and re-training of models for activities such as hyperparameter tuning have been identified as the current hot-spots [21].

The above insights emphasize that the software design is the main predictor of the environmental impacts of AI/ML systems that arise as a consequence of software compute requirements through the manufacturing and usage of hardware.

### B. Hardware Life Cycle

In the field of information and communication technologies (ICT), the dominant life cycle phases in terms of environmental impacts are typically the manufacturing (embodied impacts) and the use phase (operational impacts).

The manufacturing phase includes all raw materials, component production, electronic packaging, assembly, and testing. Existing LCAs of computers show that integrated circuits (ICs) such as processor and memory chips, as well as printed circuit boards (PCBs), connectors, and certain bulk materials are very energy and resource intensive and therefore generate a significant carbon footprint in the manufacturing phase [25]. The environmental impact of the manufacturing phase is influenced by the technology generation of the component (e.g., the miniaturization of the transistors), the production process technology, the age of the production equipment, and the available energy mix at the production site. Carbon-equivalent emissions from the production stage are also considered 'embedded emissions'.

The use phase can also have a high environmental impact depending on the processor technology, memory capacity and network configuration of the computer. In the past few years, the power consumption of the processor unit has been increasing constantly. This trend is driven by an increasing number of compute cores and higher clock frequency. Not only CPU systems but also GPU systems demonstrate this trend. High performing processor units reach a power consumption of 500 to 700 watts [26]. The environmental impact of this increasing power consumption and resulting energy density in the server racks derives primarily from the generation of electricity.

On cloud-level, key factors are the energy efficiency of the operation, including server and HVAC equipment, and the carbon intensity of the energy mix, which is determined by the share of renewable and fossil-based energy sources.

To conduct environmental assessments of the hardware component of AI/ML systems, first, a use case needs to be selected that defines the access and distribution network between the customer's end-user device and the AI server system. The AI system might be situated in a large data center thousands of kilometers away from the user. The number and type of network hops on the way to the data center should be considered in the life cycle inventory (LCI). The type and distance of the access network e.g., over cable (optical fiber) or radio (mobile communication) contributes the most to the environmental impact and needs specification [27]. Secondly, the AI systems and the related network are a shared infrastructure. Therefore, the environmental assessment of an AI service accounts for the resources and energy consumption of the physical AI system only proportionally. Due to this condition, a use pattern needs to be defined that specifies the number of users that share the system or the number and data intensity of the AI applications (services) that are created by the AI system and distributed over networks to the users. Our own calculations indicate that the network creates only a minor environmental impact in comparison to the computer system and end-user device, since it is an exceptionally large, shared infrastructure.

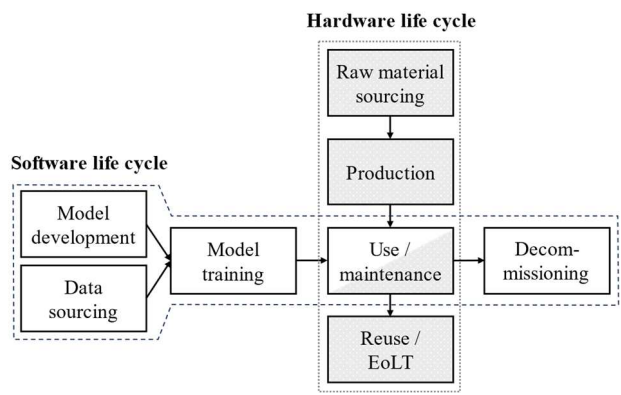

Fig. 4. Interlinkage of software and hardware life cycles

*C. Exemplary PCF of Server Hardware*

This section provides an exemplary estimation of the carbon footprint of server hardware to illustrate which technical and operational aspects of current AI/ML hardware contribute to increasing power consumption and carbon footprint. Due to data availability, a CPU-based system is discussed in more detail to demonstrate the general approach, with a following discussion on notable differences between CPU- and GPU-based systems, the latter of which are much more relevant for AI/ML systems.

The carbon footprint of the manufacturing phase of a server can be approximated based on technical data and generic conversion factors. The primary data are the type, number, manufacturing generation, and die area of the integrated circuits (ICs) [28]. A distinction must be made between CMOS technology for processors, DRAM technology for memory, and NAND for storage ICs. Another dataset includes the chip packaging technology as well as the size and number of layers of the interposer and PCB. A modern, scalable 4th generation Intel® Xeon® processor (Sapphire Rapids, 8468), which has been manufactured in Intel 7 lithography generation since 2023, serves as an example. This CPU has 48 cores and a die area of 8.12 cm² [29]. The power consumption (Thermal Design Power, TDP) is specified as 350 Watts. The CPU is suitable for the LGA4677 socket, can be equipped with up to 4 TB of DRAM and uses the Intel C741 chipset. The server mainboard used for the calculation example is the MBD X13SEM-T from Supermicro. This mainboard in Micro-ATX format has an area of 595 cm² and can be equipped with 2 TB DRAM on 8 DIMM sockets (citation needed). Regarding the 2 TB DRAM, it is assumed that a total of 1.024 modern 2 GB DRAM ICs are installed, each 0.6 cm² in die area. The remaining IC (die) area is estimated at 6 cm² based on the IC packages visible on the mainboard. The number of layers of the mainboard is not known. At least 16 layers are assumed for the example.

The carbon footprint of the production is estimated based on these data and assumptions. The manufacturing carbon footprint of the large and complex XEON CPU is estimated at 3.0 kg $CO_2$e per cm² die, the remaining, less complex ICs at 1.8 kg $CO_2$e per cm² die, and the assembled 16-layer mainboard at 0.06 kg $CO_2$e per cm² PCB. All other components and materials including connectors, direct water-cooling system, fans, power supply unit, and housing are not included in the exemplary calculation. The manufacturing of these components is less complex, and the environmental impact is mainly due to the number and volume of materials used. The carbon footprint calculation based on the above assumptions shows that the production of the CPU system amounts to almost 1,200 kg $CO_2$e (Table I). The production of the DRAM has a significant impact due to the large overall die area. If the CPU system is used for 5 years, the proportional carbon footprint per year is 240 kg $CO_2$e.

During the use phase, the power consumption of the XEON 8468 CPU is 350 W. The DRAM can be estimated at 0.2 watts per die. The DRAM therefore has an additional electrical output of 205 W. At higher clock frequency, this value can almost double. The computer system has an estimated power requirement of 555 watts in active mode. Under the assumption that this system is utilized 75% of the time with 555 W and 25% of the time in idle mode with around 200 W, the resulting annual electricity consumption is 4084 kWh. With an average carbon intensity of the energy mix of 0.4 kg $CO_2$e per kWh (typical for Germany in 2022), this results in annual emissions of 1,634 kg $CO_2$e. The total annual carbon footprint of the CPU system is therefore 1,874 kg $CO_2$e. The use phase clearly predominates with 87%. A shorter service life of the server system would influence this value. If the server were only used for three years, the annual carbon footprint would increase to 2034 kg $CO_2$e/a. If the server were operated using power from renewable energies at 0.05 kg $CO_2$e per kWh, the carbon footprint of the usage phase would be drastically reduced from 1634 kg $CO_2$e/a to 204 kg $CO_2$e/a. Under this condition the annual use phase would be slightly lower than the annual production share of 240 kg $CO_2$e/a.

As this example shows, not only the processor but also the memory configuration is influencing the environmental footprint of the server system in both the manufacturing and use phase. This general trend applies not only to CPU systems but to GPU and TPU systems as well.

*D. Exemplary Comparison of a CPU and GPU System*

The NVIDIA H100 with SXM5 board form factor is a GPU system specified with 80 GB HBM (High Bandwidth Memory) and a TDP of 700 W. The 80 GB memory configuration of the GPU system is only 2% of the 4 TB DRAM of the CPU system. At around 16 x 8 cm², the SXM5 board is significantly smaller in area than the 24.4 x 24.4 cm² Micro-ATX board of the CPU system, by over a factor four. However, the environmental assessment needs to consider the actual implementation of the H100 GPU module in an AI/ML compute system. An example is the NVIDIA DGX AI system which consists of eight H100 modules fully connected via four

TABLE I. CARBON FOOTPRINT OF SERVER COMPONENTS

| Server Components | Area [cm²] | Emission Factor | CF [kg $CO_2$e] |
|---|---|---|---|
| CPU (Sapphire Rapids 8468) | 8.1 | 3.0 | 24.4 |
| DRAM (4 TB) | 614.4 | 1.8 | 1,105.9 |
| Other ICs (e.g., chipset) | 6.0 | 1.8 | 10.8 |
| Mainboard PCB | 595.4 | 0.06 | 35.7 |
| **Total carbon footprint (CF)** | | | **1,175.8** |

NVLink switches. All the modules are mounted on a mainboard.

The significantly smaller memory capacity and compact printed circuit board results in a much lower manufacturing carbon footprint of about 200 kg $CO_2$e for the GPU system. This is a rough estimation since a bill-of-material (BOM) and other technical data such as the packaging technology of the bandwidth memory (HBM) and the PCB stack-up are not available. Moreover, the environmental assessments need to consider the entire system set-up including the power supply, network connectivity, direct water-cooling system, and housing. Nevertheless, the GPU system seems to have an environmental advantage with respect to the manufacturing phase due to the considerably lower memory capacity requirements. However, the memory capacity of GPU systems is increasing with each new system generation as well. Memory is therefore a good indicator for the environmental impact of a computer system.

Concerning the use phase, the higher average power consumption of about 700 W negatively affects the environmental footprint. Let us assume a similar use profile with 75% of the time in active mode and 25% in idle as in the case of the CPU system. The idle power consumption is unknown, and we make the assumption of 250 W. Based on these assumptions the annual power consumption accumulates to 5,147 kWh. This is 26% more power consumption per year compared to the CPU system and results at 0.4 kg $CO_2$e per kWh to 2,059 kg $CO_2$e/a. The use phase power consumption clearly dominates the environmental impact when electricity from predominantly fossil fuels is used. If renewable energy sources are used, the carbon footprint is considerably lower with 257 kg $CO_2$e/a at an energy mix of 0,05 kg $CO_2$e per kWh.

## V. METHODS FOR GREEN AI SYSTEMS

This section discusses the current status of methods for Green AI systems by discussing approaches that have been proposed in literature aiming to improve the efficiency of AI/ML systems. Following our analysis framework (Fig. 1), we first discuss approaches on software-level (model and data) and on hardware-level (server and cloud). Due to the broad scope, we do not aim to provide a comprehensive analysis, but highlight relevant and potentially impactful methods.

### A. Software: Model and Data

Methods have been researched to develop sustainability in AI/ML for years [30]. These methods have been looked at from different perspectives of AI/ML models, data, and distribution of data and models.

A commonly used approach to improve the efficiency of AI/ML models is by quantization [31], compression/pruning [32], knowledge distillation [33], and low-rank decomposition [34]. Quantization converts the high-bit floating-point parameters of AI/ML models into lower numerical precision low-bit floating-points, thus reducing the memory requirements and gaining faster computations. Methods such as sparse regularization allow pruning of redundant variables in AI/ML models during the training process, such that the overall models use less memory and obtain fast inferences. In some AI/ML models, parameters are arranged as large multidimensional arrays which can be decomposed as smaller arrays using low-rank decomposition methods. The low-rank decomposed smaller arrays can reconstruct approximations of the original parameters that use less memory and enable faster training of AI/ML models. Knowledge distillation is another useful approach that transfers the knowledge of a large, trained model to a smaller model. In practice, the development of AI/ML models by using the above methods can be more cost- and resource-efficient to obtain sustainability in AI/ML compared to improving hardware or network configurations. Their effectiveness is highly evident with many uses in popular AI/ML models. Large language models have been developed with low rank adaptation [35] to reduce the training parameters by approximately factor 10,000, requiring fewer compute budget and thereby GPUs for model training. Reference [36] have demonstrated a compression ratio by factor 71 with pruning of the convolutional neural networks used for the popular image dataset AlexNet.

Data used in modern AI/ML applications such as LLM, computer vision, and recommendation systems can be considered as big data due to both the substantial number of features and the data entries. Sustainability in AI/ML can be pursued by using data-centric methods as provided in the survey by [37], which includes methods such as data augmentation, data distillation, and few-shot learning. Moreover, dimension reduction is a well-understood and easily implementable method that can reduce the memory requirements and gain fast computations in ML/AI applications. Such dimension reduction methods can be as simple as multiplying the high-dimensional features of data with a simple randomly generated matrix to project to a small-dimensional features [38] or more complex as learning to reduce dimension by a learning models such as an auto-encoder [39]. Learning from reduced dimensions can greatly improve memory requirements and improve training times with limited amounts of loss of accuracy. Additionally, the use of appropriate optimization methods [40] for ML/AI problems is crucial for fast training times and efficient memory usages.

Apart from the above data and model-based approaches, we also want to emphasize that efficient parallelism of AI/ML models is fundamentally essential to achieve efficient computations. The advent of multi-core and multi-GPU systems with large memory have helped to achieve a significant parallelism for ML/AL models. Furthermore, popular AI programming frameworks such as PyTorch[1] and TensorFlow[2] provide support for writing code for parallel execution of AI/ML algorithms easier for programmers. Additionally, frameworks such as Colossal-AI[3] provide support on various methods of parallelism such as data parallelism, pipeline parallelism, and novel approaches such as sequence parallelism [41], zero redundancy optimizer (ZeRO) [42], and auto-parallelism [43].

Distributed learning is also a common method to achieve computational speed for tasks with high-volumes of data and computationally intensive models. One of the most investigated recent advances in distributed learning for Green AI/ML is Federated Learning [44]. This approach has a master server and distributed child nodes; however, each node learns with its own data locally and the learned parameters are communicated with the server. The server learns from the aggregate of all locally learning parameters before redistributing the updated parameters to each child node. Despite limitations of achieving sustainability targets, distributed learning and federated learning remain as an important research direction for achieving sustainability in AI.

---

[1] https://pytorch.org
[2] https://www.tensorflow.org
[3] https://hpc-ai.com

Recent research provides a summary of 30 techniques for development of Green AI/ML models, including some of the above-mentioned methods [20]. They have identified 5 tactics to investigate sustainability as data-centric, algorithmic design, model optimization, model training, and deployment. We believe that their work can be considered as a foundation for Green AI/ML to further develop practical methodologies. However, their tactics may not be complete since well-known Green AI methods such as low-rank decomposition are not considered. Additionally, AWS provides support for Green AI development[4] which can also be a valuable resource for the development of Green AI/ML.

*B. Hardware: Server and Cloud*

While the discourse on Green AI/ML is still relatively new, the discussion and implementation of strategies to minimize environmental impacts of computer hardware and data centers has been ongoing for decades (e.g., [45][46]). We therefore do not aim to provide a comprehensive overview of individual measures, but discuss select strategies and initiatives of higher perceived relevancy.

In 2021, cloud infrastructure providers and data center operators in Europe formed the 'Climate Neutral Data Center Pact', a self-regulatory initiative to make European data centers climate neutral by 2030 via measures including the usage of low-carbon energy, water conservation, heat recycling, and reuse and repair of hardware [47]. In the US, Google has pledged to run of their global data centers on low-carbon power by 2030. Microsoft has declared to be carbon-negative by 2030 and offset all their historical carbon footprint by 2050 [48]. Two main ways have been described to achieve carbon-neutral data centers in practice: First, by maximizing energy efficiency and the use of low-carbon energy, and second, by carbon offsetting [49].

Optimizing the energy consumption of cloud computing has long been a topic of interest, both due to economic considerations but also environmental concern. For instance, load balancing is conducted to distribute the workload across multiple nodes in a shared pool of computing resources. It helps in optimal utilization of available resources and enhancing the performance of the system, thereby reducing energy consumption and the associated carbon emission rate (e.g., [50]). Regarding HVAC, Google reported already several years ago to have identified a method to reduce data center cooling needs by 40%, by training a deep learning model on historical data and predicting conditions in the immediate future to increase the overall efficiency [51].

Beyond minimizing energy consumption of data centers, reducing the carbon intensity of consumed energy is key, such as by locating data centers in close proximity to renewable energy sources, such as wind parks. Such measures are of high relevance as the power consumption of AI systems dominates their overall carbon footprint with approximately 80% of their total emissions, as we have highlighted in section IV.

While the above strategies and initiatives are mostly concerned with reducing the operational impacts of data centers, minimizing the embedded impacts of server and cloud hardware is also key for the realization of Green AI. In the EU, Regulation (EU) 2019/424 on ecodesign requirements for servers and data storage products [52] under the European Ecodesign Directive (2009/125/EC) sets minimum requirements to limit the environmental impact of servers and data storage products. While focusing primarily on energy efficiency, the regulation also includes material efficiency requirements, including the use of joining, fastening or sealing techniques that do not prevent the disassembly of several key components, including CPU, GPU, memory, and mainboard to enhance repair and reuse, and further mandates secure data deletion techniques to facilitate reuse.

Beyond such measures, resource efficiency during the fabrication of server equipment (embedded impacts) need to be addressed. Naturally, maximizing the use of low-carbon energy in the fabrication facilities for semiconductors, PCBs, and other required server hardware that powers AI/ML is an effective lever to minimize embedded carbon. Further, striving to apply principles of the circular economy, including lifetime extension of the hardware including through reuse, repair, and remanufacturing, maximizing collection and material recovery rates in recycling processes can all be effective measures. This is already practiced, at least in part, with for instance Amazon Web Services (AWS) having implemented refurbishment processes for data center hardware to extend its useful life [53]. Other companies like Google and Microsoft have also introduced circular and zero-waste strategies to boost reuse and recycling of hardware to maximize resource efficiency and avoid emissions from the production of new equipment [53].

As a last example, optimizing hardware and matching of software requirements and hardware capabilities is used to contribute to efficiency and thereby Green AI. While GPU are more proficient than CPU for typical AI/ML workloads, more specific and dedicated hardware like TPU, ASIC (application-specific integrated circuits) and FPGA (field-programmable gate array) can be more efficient for specialized tasks, increasing the achievable operations per unit of input energy.

## VI. DISCUSSION OF FUTURE RESEARCH

This section discusses what we believe is missing from the current discourse on Green AI and/or where future research could contribute to its further development and uptake. We make proposals for future research on software-level (model and data), hardware level (server and cloud), and related to environmental assessment and metrics.

*A. Proposals on Software-Level: Model and Data*

The most effective method to achieve sustainability in AI/ML is by designing AI/ML models to minimize the resource usage and the carbon footprint. Therefore, it is important to further research on sustainable methods on AI/ML model design, data processing, optimization, deployment, and distributed learning. A major practical limitation in developing Green AI models (discussed in Section V) is the technological complexity of implementing methods such as quantization, pruning, knowledge distillation, and low-rank decomposition. Incorporation of such methods to existing AI/ML systems may require a considerable amount of expertise, time, and resources, which are not always at hand for companies, institutes, and individuals. Hence, we want to propose several strategies to make Green AI models development more practical.

---
[4] https://aws.amazon.com/blogs/architecture/optimize-ai-ml-workloads-for-sustainability-part-1-identify-business-goals-validate-ml-use-and-process-data/

Awareness and availability of knowledge and resources of Green AI methodology for model developers is one of the primary and effective ways of making wide-spread usage of Green AI models. We believe the development and practice of (Machine Learning Operations) MLOps and design patterns [54] for Green AI is vital. Amazon Web Services (AWS) already provides support for MLOps for sustainable AI. Furthermore, use of software frameworks such as Colossal-AI that can leverage the burden of parallelism of AI/ML models from software engineers and researchers are useful to make sustainability both cost- and time-efficient. Additionally, we emphasize that a movement towards free and open-source software frameworks to assist in the development of Green AI/ML models can increase the wide-spread usage of sustainable practices.

Availability of tools and programming language/ framework support for Green AI/ML is also conducive towards sustainable AI practices. In addition to the support for parallelism, AI programming frameworks should provide tools and support to implement sustainable methods such as pruning, low-rank decomposition and knowledge distillation. Availability of easy-to-use programming libraries for sustainability support with popular AI frameworks such as PyTorch and TensorFlow can be highly impactful. At a higher level of research, it should also consider AI/ML itself learning to convert AI/ML code into code with sustainable features based on quantization, knowledge distillation, low-rank constructions, pruning, automatic parallelism, and processing of data. We may call such a method "AI4greenAI", however, we believe that this is a challenging task which may require a considerable amount of research. Emerging fields of automatic code generation [55] and theorem proving [56] based on LLM may serve as foundation in research for AI automatically generating code for Green AI. Ongoing research on detection and generation of parallel code using machine learning for code augmentation and compiler optimization [57][58][59] are motivations to believe that the same should be possible for Green AI.

We also believe more research should be conducted in Green AI model development methods. Research on less utilized methods such as random projection [60] and random Fourier features [61] may have the potential to benefit sustainability due to their ability to learn with small memory requirement and fast computation [62]. Another important direction is the development of AI/ML models with constraints of sustainability built into the model. It would be useful to investigate whether it is possible for an AI/ML model to achieve learning training objectives (e.g., prediction accuracy) and sustainability constraints simultaneously. Some research in this direction has already been conducted by [63], where they propose pruning of weights of neural networks with predefined budget on Multiply-and-Accumulate (MAC) operations. One of the future research directions would be to further identify sustainability constraints that can be quantified and integrated in AI models. Sustainability constraints can be problems dependent or common constraints such as amount of memory usage and amount of network communications. Furthermore, sustainability constraints could be incorporated with sustainable model design methods such as low-rank decomposition and knowledge distillation. Extended research should also consider self-adaptation of the learning models towards periodically changing energy and $CO_2$ emission capacity. There can also be competitive resource sharing among different AI models to optimize objectives with maximum utility and sustainability constraints modeled (i.e., game theoretic frameworks [64]).

In addition to the above, more research is needed on design of data centers, hardware, and communication systems together with federated learning (and distributed learning, in general) to achieve practical solutions with sustainability. We also believe that further research on load balancing [65] and resource adaptive models [66] are highly necessary. Compressed communication for distributed DL/ML [67] is another research direction that can greatly contribute towards Green AI/ML. Furthermore, there can be coherence in the development of electronics and hardware systems to adhere to theoretical understanding obtained with distributed machine learning models such as federated learning. Initiatives such as TinyML[5] can greatly influence the future development of sustainable hardware design along with algorithmic development of AI/ML models.

We acknowledge that there may be more paths on the software-level towards Green AI, such as approximate computing [68] just to name one, but the breadth of the subjects limits the number of examples included in our discussion.

### B. Proposals on Hardware-Level: Server and Cloud

As highlighted prior, the state of knowledge on ecodesign measures for hardware appears, to us, more developed compared to the software level, not last due to the more recent and dynamic nature of the technological progress in AI/ML software. Nevertheless, more work needs to be done on hardware-level to achieve Green AI. In general, hardware needs to be selected and configured for the specific application purpose. Hardware includes computing, storing and network functionality and needs to be holistically optimized. The hardware should also be load adaptive and utilized as best as possible.

Further research could contribute to maximizing the energy efficiency, or mathematical operations per watt of energy, of the hardware. Specialized accelerators for AI have the potential to further reduce the power consumption of neural networks, including analog accelerators, a category of devices utilizing non-volatile memory to store weights and perform in-situ analog computation [69]. Compute-in-memory (CIM) based on resistive random-access memory (RRAM) has been described to achieve substantial energy efficiency gains by storing AI model weights and performing computations directly in RRAM devices, thereby eliminating energy-consuming data movement between separate compute and memory components [70]. Such approaches further enable increasingly complex AI/ML models to operate

---

[5] https://www.tinyml.org/

directly on edge devices [70], eliminating data traffic between data centers and periphery.

Within the AI data centers liquid cooling (direct chip cooling) is becoming the norm. This technology allows better thermal energy transportation on the one hand and improved reuse of this energy on the other, as carbon intensive generation of heat will be avoided. In order to utilize this energy saving potential, research and product development should focus on developing technologies for the lossless transport of thermal energy (exergy) from the chip to the room/infrastructure level and its warm water transport system.

To reduce the total energy consumption, the highest possible utilization of the existing hardware, for example through virtualization, must be striven for. Also, the equipment con-figuration should support as best as possible the specific application of the server system. In order to obtain the optimal system configuration (e.g. CPU core count, clock frequency, L3 cache, memory bandwidth and capacity), energy efficiency benchmarks (e.g. SPEC SERT) should be used and further developed.

The application of reinforcement learning for task scheduling in a federated cloud environment has been suggested to select the data center for individual tasks based on factors such as energy consumption, cooling method, waiting time of the task, energy type, emission ratio, and total energy consumption of the cloud data center [73]. While this may indeed be an effective method to lower the carbon footprint of AI applications, in order to realize net benefits, additional capacity for the generation of low-carbon energy needs to be installed proportionately to the energy consumption of the AI system. Otherwise, the total amount of available low-carbon energy in a geographical region remains unchanged and is simply diverted from other applications.

Referring to the classical 3R principles of ecodesign, to improve the material efficiency on cloud-level, a relevant lever is to identify the optimal hardware refresh cycle that considers both the embedded impacts of the hardware as well as potential increases in energy efficiency of more recent hardware, which can reduce the overall need for production of new equipment. Once decommissioned, following the principles of the circular economy, reuse should be prioritized over recycling to maximize the utility derived from the hardware and its embedded environmental impacts. Finally, more research is needed on recycling processes that maximize the recovery rates for as many of the materials embedded in the hardware as possible, ideally at a level of quality that enables closed-loop integration of recyclates into the production of new equipment.

*C. Proposals on Assessment and Metrics-Level*

Quantification enables the verification of effective measures. Suitable assessment methods, metrics, and benchmarking are essential to track relevant indicators such as energy use, carbon emissions, and resource consumption. However, challenges remain, some of which are addressed in this section.

For the development of Green AI, the development and integration of analysis and prediction tools for compute, energy use, and carbon footprint into software development environments are key. First promising implementations are already available, such as CodeCarbon[6], a Python package that enables developers to estimate the carbon footprint of their work, measuring the power usage of utilized hardware (CPU, GPU, RAM) combined with estimations on the carbon intensity of the consumed energy by cloud provider and geographical region. However, currently, the integration of life cycle emissions from the computing infrastructure (hardware) is still lacking. More work on such practicable tools could drastically reduce the barrier for AI/ML researchers and engineers to account for sustainability considerations from the early phases of model development. Further, for companies and researchers to routinely communicate key data on the energy efficiency and environmental performance of their models could be a first step towards increasing transparency and a race to the top for researchers and companies to achieve Green AI models.

Another challenge for impact assessments of AI/ML systems is that life cycle data on key processes and components is severely lacking, particularly on semiconductor fabrication processes and devices. We consider research and industry participation is needed to enable more reliable and timely estimates of the environmental impacts of hardware used for AI/ML systems.

Although energy consumption and the carbon footprint are the most obvious environmental challenges associated with AI systems today, we highlight that other indicators such as the resource consumption and water footprint also require attention. It is well known that electronics contain numerous critical raw materials, including chip-grade silicon, boron, arsenic, phosphorus, gallium, and copper, among others, and that the recovery rates in current end-of-life treatment processes remains low [72]. Other indicators, such as the water footprint, may also require more research to enable accurate assessments [73].

Further research is also needed for remaining methodological challenges related to environmental assessments of AI/ML systems. For instance, defining an appropriate functional unit remains challenging. It requires correlating the utility of the AI system (the service) to its manufacturing and use related environmental impact. For an AI model that generates images, an appropriate FU may be the carbon footprint of one generated image. However, the FU should also account for qualitative aspects such as the resolution or complexity of an image as well as the time it takes for creating the image. In this context, we consider that in conjunction with research activities, standardization efforts could play a vital role in achieving better metrics. On a European level, CEN-CENELEC has created the CEN/CLC/JTC 21 on Artificial Intelligence, for which one theme is green and sustainable AI. The BSI website lists a document in preparation, aiming to establish a framework for quantification of environmental impact of AI and its long-term sustainability, and to encourage AI developers and users to improve efficiency of AI use.

VII. SUMMARY AND OUTLOOK

Given the potential for AI/ML systems to incur substantial environmental impacts, with this paper, we aim to broaden the scientific discourse on Green AI and to raise awareness among researchers and stakeholders in the field of AI / ML as well as the hardware and data center industry. We consider that system thinking is needed to account for the life cycle impacts of both software and hardware systems and have proposed a supporting framework to support such analyses.

---

[6] https://codecarbon.io/

For the path towards Green AI, we propose that first, raising awareness among stakeholders from industry and academia is needed to strengthen the development of Green AI methods and to boost uptake of its principles in practice. Second, practical tools that measure and predict the computational requirements, energy consumption, and environmental impacts of AI/ML models need to be integrated into all phases of model development and data management. This may include the adaptation of the classical "3R" principles of reduce, reuse, recycle, to software systems. Third, given the complexity and breadth of measures that have been suggested in literature, to lower the barrier of their application in practice, research should focus on leveraging AI to automate mechanisms for Green AI both during software development and during use – a vision we refer to as 'AI4greenAI'.

We acknowledge the different timescales for the pace at which the discussed factors for Green AI – software, hardware, and energy – can be addressed. There is a considerable focus on the development of ML models and AI-based systems, making it a highly dynamic and fast-paced field. Development and implementation of methods to reduce the energy consumption of AI models can therefore occur at a comparatively higher pace. In contrast, although the development of hardware for AI applications is also proceeding at a high pace, the timescale for energy efficiency gains is comparatively longer, due to longer development cycles from conceptualization to fabrication of semiconductors and servers. Lastly, while essential, the timeline for increasing the capacity of low-carbon energy systems may be longer than both software and hardware development. Nevertheless, efforts on all three levels – software, hardware, and energy – are needed in parallel on the path towards Green AI.

ACKNOWLEDGMENT

This work has not received any third-party funding.